\documentclass[pre,letterpaper,twocolumn,floatfix,superscriptaddress]{revtex4}
\usepackage{graphicx}

\begin{document}

\title{Fast Fourier Transform Simulation Techniques for Coulomb Gases}

\author{A.~Duncan}
\affiliation{Department of Physics and Astronomy\\
University of Pittsburgh, Pittsburgh, PA 15260}
\author{R.D.~Sedgewick}
\altaffiliation[Current address: ]%
{Department of Biological Sciences \\
Carnegie Mellon University \\
Pittsburgh, PA 15213}
\affiliation{Department of Chemistry\\
University of Pittsburgh, Pittsburgh, PA 15260}
\author{R.D.~Coalson}
\affiliation{Department of Chemistry\\
University of Pittsburgh, Pittsburgh, PA 15260}


\begin{abstract}

An improved approach to updating the electric field in simulations of
Coulomb gases using the local lattice technique introduced by Maggs
and Rossetto \cite{maggs:prl}, is described and tested. Using the Fast
Fourier Transform (FFT) an independent configuration of electric
fields subject to Gauss' law constraint can be generated in a single
update step.  This FFT based method is shown to outperform previous
approaches to updating the electric field in the simulation of a basic
test problem in electrostatics of strongly correlated systems.

\end{abstract}
\maketitle
\section{Introduction}

Electrostatic forces play a major role in intra- and inter-molecular
interactions.  In particular, in locally charged systems, strong
Coulomb interactions between charged particles often dominate
structural and dynamical properties \cite{rob1} . This situation
occurs frequently in biological systems, where both protein
\cite{rob2,rob3} and DNA macromolecules \cite{rob4} typically undergo
H+ dissociation or association, thereby generating a charged macroion
with counterions in solution.  In addition, aqueous solutions often
contain dissolved salts, thus introducing more mobile ions into the
system.  The thermally averaged interaction force (which integrates to
give the ``potential of mean force'' (PMF) \cite{rob5, rob6}) between
two macroions depends strongly on electrostatic forces that are
supplied by the mobile ions in the surrounding solution.  Calculation
of the PMF, or, equivalently, electrostatic free energy of interaction
between two macroions, is critical to understanding their equilibrium
(and, ultimately, dynamical) properties.  For example, the
electrostatic PMF determines the binding equilibrium constant for the
macroion pair \cite{rob7}. Similar remarks hold for synthetic
macroions, for example, charged colloid particles, whose self-assembly
properties, phase diagrams, etc., are determined by precisely the same
type of considerations \cite{rob8, rob9}.  Dynamical properties, such
as rate constants for transitions between stable states \cite{rob6}
and ion currents through channel proteins \cite{rob10}, are heavily
influenced by the energy landscape determined from equilibrium free
energy considerations.


The difficulty of implementing such long
range (``nonlocal'') forces in realistic numerical simulations is well
known: each charged particle ``feels'' the effect of all the others,
so that naive algorithms typically scale with the square of the system
volume $V$, while more sophisticated algorithms (Ewald summation,
Fourier techniques \cite{ewalda, ewaldb,fft}) with better scaling properties
nevertheless prove far from efficient in computing the electrostatic
energy for large systems. Recently, a local formulation of the Coulomb
gas problem (first introduced by Maggs and collaborators \cite{maggs:prl} has
offered a possible exit from this impasse. At the cost of including an
unphysical transverse (or ``curl") part in the electric field, the
effects of which on the charged particle dynamics can be shown to
decouple on average over a suitably long Monte Carlo simulation of the
system, the Hamiltonian of the system can be recast in a completely
local form, so that the computational cost of a system update becomes
of the order of the system volume $V$ (rather than $V^2$, as
above). This method, together with several improvements designed to
improve the mobility of the charged particles in the simulations, have
been the subject of a number of recent publications \cite{maggs:prl, maggs:worm, maggs:trail,ourpaper}.

The local Coulomb gas formalism is based on 
the partition function of a system consisting of $N$ charges (mobile or fixed)
$e_i$ at locations $\vec{r}_{i}$, thereby producing a charge density
\begin{equation}
\label{eq:chargedens}
   \rho(\vec{r}) = \sum_{i}e_{i}\delta(\vec{r}-\vec{r}_{i})
\end{equation}
Accordingly, the canonical partition function for the system at
inverse temperature $\beta$ becomes
\begin{equation}
\label{eq:partfunc}
  Z=\int\prod_{i=1}^{N}d\vec{r}_{i}{\cal D}\vec{E}(\vec{r})\prod_{\vec{r}}\delta(\vec{\nabla}
\cdot\vec{E}-\frac{4\pi}{\epsilon}\rho(\vec{r}))e^{-\frac{\beta\epsilon}{8\pi}\int d\vec{r}\vec{E}^{2}
(\vec{r})}
\end{equation}
where $\epsilon$ is the dielectric constant of the medium in which the
charges move (assumed spatially uniform throughout this paper). The
essential point is that the transverse, or curl, part of the electric
field variable decouples from the charged particle dynamics via the
Helmholtz decomposition
\begin{eqnarray}
\label{eq: Helm}
  \vec{E}&=& \vec{\nabla}\phi + \vec{\nabla}\times\vec{A}  \\ 
  \int d\vec{r}\,\vec{E}^{2} &=& \int d\vec{r}\, |\vec{\nabla}\phi|^{2}+\int d\vec{r}\,|\vec{\nabla}\times\vec{A}|^{2}
\end{eqnarray}
As only the gradient part of the electric field appears in the Gauss'
Law constraint in (\ref{eq:partfunc}), the curl part clearly decouples
from the particle positions. However, the simulation of the integral
in (\ref{eq:partfunc}) involves charged particle moves that are only
sensitive to the local value of the electric field, provided that the
local field is modified to maintain Gauss' Law at all
times. Practically, the simulations are most readily performed on
spatial lattices, with the particle charges associated with lattice
sites, and the electric fields with links connecting nearest neighbors
on the lattice. For a complete description of the algorithm, the
reader is referred to Ref.~\cite{maggs:prl}.

\section{Previous Work}

Before discussing our FFT based approach to updating the electric
field in a local Coulomb gas simulation, we briefly review some of the
field update methods that have been used in previous applications of
the local lattice method of Maggs et al \cite{maggs:prl, maggs:worm, ourpaper}.

The implementation of the method is most simply seen on a square
lattice with dimensionless variables.  We therefore scale the electric
field by a factor of $\epsilon a^2/4 \pi e$ and the inverse
temperature by a factor of $4 \pi e^2/\epsilon a$, where
$\epsilon$ is the dielectric constant of the system and $a$ is the
lattice spacing.

Any field update method must conserve Gauss' law at every step of the
simulation.  The simplest methods for updating the electric field are
based on shifting the four link fields of a plaquette by the same amount
$\alpha$.  Such a shift maintains the Gauss' law constraint.  One
possibility is to use a
Metropolis procedure to pick $\alpha$, but it is more efficient to use
a heat-bath method.  Considering one plaquette and labeling its links
1 through 4, the portion of the Hamiltonian that depends on
these links (written using dimensionless variables) is given by
\begin{equation}
H_p = \frac{1}{2} \sum_{i=1}^4 (E_i + \alpha)^2
\end{equation}
where $E_i$ is the electric field on the $i$th link in the direction
obtained by going around the plaquette clockwise. This can be rewritten as
\begin{equation}
H_p = \frac{1}{2} (\alpha - \sum_i E_i)^2
\end{equation}
where we have dropped terms with no $\alpha$ dependence.  Therefore $\alpha$
can be generated with the canonical distribution by setting it to
$x/\sqrt{ \hat{\beta}} - \sum_i E_i$ where $x$ is a Gaussian-distributed variable with zero mean and standard deviation one and
$\hat{\beta}$ is the dimensionless inverse temperature.  Updating the
plaquettes using this procedure will be referred to in this paper as the local
heat bath method, as the change in each plaquette is affected only by the
immediate local neighborhood of that plaquette. Of the three methods compared
in this paper, the local heat bath is the simplest to implement but also the
``most local," and can be expected to yield the longest autocorrelation times.

Level, Alet, Rottler and Maggs \cite{maggs:worm} recently introduced a
cluster move to update the electric fields more efficiently based on a
worm algorithm developed by Alet and S\o rensen \cite{alet:orig_worm}.
The method introduces a pair of positive and negatively charged
particles on a randomly chosen site of the lattice.  One of the
particles then moves about the lattice in a biased random walk, as
described in Ref.~\cite{alet:orig_worm}.  This pseudo-particle moves
through the lattice (modifying appropriately the electric field to
preserve Gauss' law), eventually returning to its partner where the
two annihilate each other.  A loop of links with modified electric
fields is left behind along the path taken by the mobile particle.
This change of the electric fields along this loop is then accepted or
rejected depending on the fields entering the initial site.
Empirically these loop changes are accepted with high probability.
The charge of the virtual pair is randomly chosen with a uniform
distribution in the interval ($-e_m$, $e_m$), where $e_m$ is a free
parameter which is chosen to maximize efficiency.

The worm method is clearly more ``global" than the local heat bath
method described above--- typically the number of links modified is of
order the lattice volume--- and we may expect autocorrelation times
that are shorter than those obtained with plaquette
updates. Nevertheless, field configurations that differ by a single
worm update are still substantially correlated with each other.  Next,
we show that a global regeneration of the electric field is possible,
using FFT techniques, which is computationally more efficient than the
worm method on reasonably large lattices, and produces a completely
decorrelated electric field after a single update step.

\section{Fast Fourier Decorrelation Method}

In this section we describe in some detail a FFT-based approach to
updating the electric field in local Coulomb gas simulations that is
computationally of order $V \log\ V$, and produces a globally
decorrelated transverse field at each update. It is similar to
methods used to study electromagnetic splittings in lattice quantum
chromo-dynamics in Ref.~\cite{tonyqcd}. The method does not have
any free parameters, obviating the need to optimize simulation
parameters, and, as we shall see below in some sample computations, is
computationally more efficient (in the sense of computational effort
per autocorrelation time of physically interesting observables) than
the other updating schemes described previously.  

  Once again, consider a lattice Coulomb gas with charge density $\rho_{n}$
and electric field link variables $E_{n\mu}$ (where $n$ denotes lattice
sites and $\mu=1,2,3$ spatial direction). The partition function takes
the form, for fixed charge particle locations (i.e. fixed $\rho$),
\begin{equation}
\label{eq:boltzwt}
  Z=\int dE_{n\mu}\delta(\bar{\Delta}_{\mu}E_{n\mu}-\rho_{n})e^{-\frac{\hat{\beta}}{2}\sum_{n\mu}E_{n\mu}^{2}},
\end{equation}
where $\bar{\Delta}$ (resp. $\Delta$)  denote left (resp. right) lattice
derivatives here and below.  We wish to develop an efficient procedure for
traversing the space of electric field configurations $\{E_{n\mu}\}$
consistent with the Gauss' Law constraint. A perfect decorrelation can be
achieved if the new field has a transverse (``curl'') part completely
decorrelated with the previous transverse field: on the other hand, the
longitudinal (``gradient'') part of the field is fixed by the Gauss' Law
constraint and must be preserved by the update procedure. To accomplish this
we must (a) extract the longitudinal part, and (b) generate a new, and
completely independent transverse part of the field. 

We shall work on a $L$x$L$x$L$ lattice (volume $V=L^3$) with 
Fourier transform fields defined as follows
\begin{eqnarray}
  \rho_{n} &=& \frac{1}{V}\sum_{k} e^{ik\cdot n}\rho(k)  \\
  E_{n\mu} &=& \frac{1}{V}\sum_{k} e^{ik\cdot n}E_{\mu}(k)
\end{eqnarray}
Note that we distinguish lattice coordinate space fields from their momentum
space transforms by using subscripts for the former and function notation for
the latter. Up to constant field configurations (which are separately
simulated as explained in Ref.~\cite{maggs:prl}), an arbitrary lattice vector field $E_{n\mu}$
may be decomposed
\begin{eqnarray}
\label{eq:etot}
  E_{n\mu} &=& E^{||}_{n\mu}+E^{\mbox{tr}}_{n\mu}  \\
\label{eq:epar}
  \bar{\Delta}_{\mu}E^{||}_{n\mu} &=& \rho_{n} \\
\label{eq:etrans}
  E^{\mbox{tr}}_{n\mu} &=& \epsilon_{\mu\nu\rho}\bar{\Delta}_{\nu}A_{n\rho}
\end{eqnarray}
where $A_{n\rho}$ is a transverse lattice vector field satisfying
$\Delta_{\rho}A_{n\rho}=0$ (as we see from Eq.~\ref{eq:etrans}, the gradient
part of $A_{n\rho}$ is absent from $E^{\mbox{tr}}$). Consequently, the
Fourier transform of $A_{n\rho}$, defined through
\begin{equation}
   A_{n\rho}=\frac{1}{V}\sum_{k} e^{ik\cdot n}A_{\rho}(k)
\end{equation}
must satisfy
\begin{equation}
   s^{*}_{\rho}A_{\rho}(k)=0,\;\;\;s_{\rho}\equiv 1-e^{-ik_{\rho}}
\end{equation}
and can accordingly be written in terms of polarization vectors
$\vec{\epsilon}_{1}, \vec{\epsilon}_{2}$ as follows
\begin{eqnarray}
\label{eq:afour}
   A_{\rho}(k) &=& a_{1}(k)\epsilon_{1\rho}(k)+a_{2}(k)\epsilon_{2\rho}(k)  \\
\label{eq:eps1}
   \vec{s}^{*}\cdot \vec{\epsilon}_{1}&=&\vec{s}^{*}\cdot \vec{\epsilon}_{2}=0 \\
\label{eq:eps2}
   \vec{\epsilon}_{1}\cdot\vec{\epsilon}^{*}_{2} &=& 0
\end{eqnarray}
for a discrete mode $k$ corresponding to a complex Fourier component (i.e.
where $k_{\mu}=2\pi n_{\mu}/L$ with not all of $n_1,n_2,n_3$ equal to 0 or
$L/2$). For the real modes the reader may easily verify that analogous
formulas hold involving purely real vectors. From the form of the partition
function given in Eq.~\ref{eq:boltzwt} one finds that the Fourier coefficients
$a_{1},a_{2}$ in Eq.~\ref{eq:afour} are to be generated according to the
Gaussian weight
\begin{equation}
\label{eq:gausswt}
  Z(a_1,a_2) = e^{-\frac{\hat{\beta}}{V}\Delta(k)(|a_{1}|^{2}+|a_{2}|^{2})}
\end{equation}
with $\Delta(k)\equiv 4\sum_{\rho}\sin^{2}{(k_{\rho})}$. Once random
polarization vectors $\vec{\epsilon}_{1}$ and $\vec{\epsilon}_{2}$ are
generated satisfying Equations \ref{eq:eps1} and \ref{eq:eps2}, then
Eq.~\ref{eq:afour} yields the complete Fourier transform $A_{\rho}(k)$. An
inverse FFT then yields the coordinate space field $A_{n\rho}$, and a lattice
curl the desired transverse part of the electric field $E^{\mbox{tr}}$ via
Eq.~\ref{eq:etrans}. 

This heat bath procedure clearly produces a new transverse electric field
completely decorrelated from the preceding one. As we wish to update the
\textit{total} electric field, we must also calculate the gradient part
$E^{||}$, which can then be added to the new decorrelated transverse field.
To do this, we simply note that from Eq.~\ref{eq:epar} it follows that 
\begin{equation}
  E^{||}_{\rho}(k) = s^{*}_{\rho}(k)\frac{\rho(k)}{\Delta(k)}
\end{equation}
where $\rho(k)$ is the Fourier transform of the charge density $\rho_{n}$,
which is also to be computed by FFT. Thus this algorithm involves (apart from
the effort required to generate the Fourier components $A_{\rho}(k)$ as
indicated above) 4 FFT operations to generate a globally decorrelated electric
field.  As we shall see below, this computational effort compares favorably
with competitive methods, such as local heat-bath or worm type update
algorithms. 

\section{Results}

To test our FFT-based method for updating the electric fields, and compare its
efficiency with that of the local heat-bath and worm update methods, we
simulated the system of charged conducting plates with ions between the plates
discussed in Ref.~\cite{ourpaper}.  The basic system consists of a 50x50x50
lattice with periodic boundary conditions in all three dimensions.  Positive
charges are free to move on two fixed plates separated in the $x$ direction
that extend the entire extent of the lattice in the $y$ and $z$ directions,
while the region between the plates contains mobile counterions ensuring
overall neutrality.  As in Ref.~\cite{ourpaper}, the lattice spacing is chosen
to be 1 \AA, the dielectric constant is $80.0$, and the temperature is $300 K$
so that the dimensionless inverse temperature, $\hat{\beta}$, is $87.1$.  The
plates are charged with 34 positively charged univalent ions, which are
constrained to stay on the plate.  34 negatively charged divalent ions are
placed between the plates, and are excluded from the plane of lattice sites
closest to the plates.  Every lattice site can be occupied by at most one ion.
The ions are moved by the coupled particle-field heat bath method discussed in
Ref.~\cite{ourpaper}.

Each Monte Carlo step is composed of (200 $\times$ number of charges
on the plates) attempted moves of the positive ions on the plates,
(2000 $\times$ number of ions in solution) attempted moves of negative
ions in solution, a global update of the total electric field (as
described in Ref.~\cite{maggs:prl}), and an update of the electric
field using either the local heat bath method, the worm method, or the
FFT based method.  For the local heat bath method $3 \times 50^3$
links are chosen at random to be updated. For the worm method, 5 worms
are created for each Monte Carlo sweep.  The maximum ghost charge used
in the worm method was set to 0.3, which was determined experimentally
to be the value that minimized the autocorrelation times of our
observables.  For the FFT method the FFTW package \cite{fftw} was used
to perform the FFTs.

On a 2GHz AMD processor it takes 336 seconds to perform 1000 updates of the
electric field using the FFT method.  It takes 432 seconds to perform 1000
Monte Carlo sweeps using the local heat bath method.  It takes 700 seconds to
perform 1000 Monte Carlo sweeps using the worm method, where each Monte Carlo
sweep corresponds to creating 5 worms.  Thus, using these parameters, the FFT-based method is more than a factor of 2 faster than the worm-based method.  

To compare the efficiencies of the different methods, we consider the
autocorrelation times of various observables.  The autocorrelation of an
observable $A$ is given by
\begin{equation}
C(t) = \frac{\sum_i (A_i -\bar{A})(A_{i+t} -\bar{A})}
            {\sum_i (A_i - \bar{A})^2},
\end{equation}
where $A_i$ designates the $i$th measurement of $A$ and $\bar{A}$ is the
average value of $A$.  We extract the autocorrelation time of an observable,
$\tau$, by integrating the autocorrelation function out to a distance where the measurements are decorrelated ($\tau = \int_{t=0}^{t_{max}} C(t)$ where $C(t_{max}) \approx 0$).


\begin{figure}
\centerline{\includegraphics[width=3.5in]{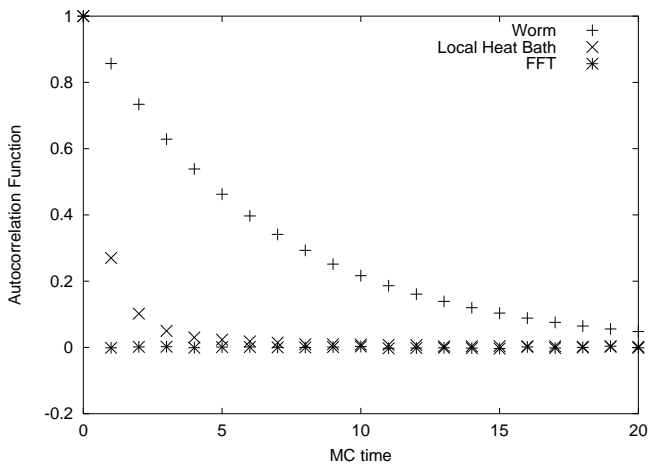}}
\caption{ Autocorrelation function of the electric field in the $y$
  direction at the (25,25,25) lattice site from simulations using
  FFT-based update method, local heat bath update method, and worm
  update method.  After equilibration the particles are frozen in
  place for these runs.}
\label{fig:efield_autocorr}
\end{figure}

Initially we consider the autocorrelation of the electric field on a
single link of the lattice.  In order to observe how the electric
field decorrelates itself using solely the different field update
methods, we fix the location of the ions after the warmup sweeps.  We
then simulate the electric field using the different field update
methods and observe the electric field in the $y$ direction on the
(25,25,25) site.  The autocorrelation function of this electric field
is shown in Fig.~\ref{fig:efield_autocorr}.  We determine the
autocorrelation time by integrating the autocorrelation function out
to 40.  This gives an autocorrelation time of $7.04$ for the worm
method and an autocorrelation time of $1.61$ for the local heat bath
method.  The FFT based method gives a completely decorrelated set of
electric fields, so the electric field under observation is
decorrelated after a single Monte Carlo step and the autcorrelation
time is $1$.  The local heat bath method is able to decorrelate this
observable rapidly because the observable is so local.

\begin{figure}
\centerline{\includegraphics[width=3.5in]{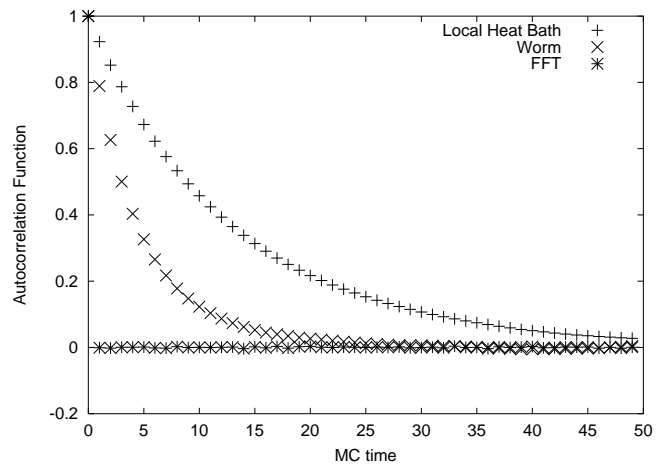}}
\caption{ Autocorrelation function of the (2, 2, 2) component of the
  Fourier Transform of the electric field in the $x$ direction from
  simulations using FFT-based update method, local heat bath update
  method, and worm update method.  After equilibration the particles
  are frozen in place for these runs.}
\label{fig:fft_autocorr}
\end{figure}

We next consider the autocorrelation function of the (2, 2, 2)
component of the Fourier Transform of the electric field in the $x$
direction.  Again we fix the location of the ions to concentrate on
the efficiency of the field update methods.  The autocorrelation
functions are shown in Fig.~\ref{fig:fft_autocorr}.  Integrating the
autocorrelation function out to 50 gives an autocorrelation time of
$13.3$ for the local heat bath update method runs and $5.24$ for the
worm update method runs.  The local heat bath method has more
difficulties with smaller momentum components of the
Fourier-transformed electric field: the (0, 0, 1) component has an
autocorrelation time of $123.3$ with the local heat bath update, while
the autocorrelation time with the worm update is only $3.36$.  For
this observable the large clusters that are updated with the worm
method allow it to decorrelate these small-momentum Fourier components
more rapidly then the local heat bath method.  The FFT based method
decorrelates all the Fourier components of the electric field in a
single Monte Carlo update step.


\begin{figure}
\centerline{\includegraphics[width=3.5in]{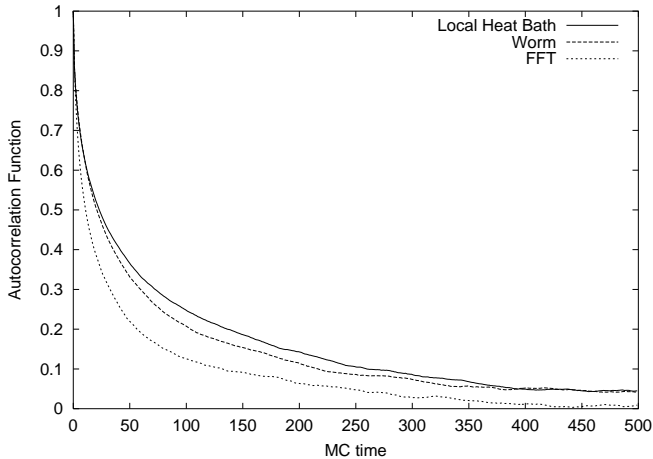}}
\caption{Autocorrelation function of the particle-particle energy from
  simulations using FFT based update method, local heat bath update
  method, and worm update method.}
\label{fig:ppe_autocorr}
\end{figure}

We then perform the simulations with the ions allowed to move throughout 
the simulation and consider the
particle-particle energy given by the expression 
\begin{equation}
E_{pp} = \frac{1}{4 \pi} \sum_{i}^{N_{part}} \sum_j^{j<i} \frac{q_i q_j}{r_{ij}}.
\end{equation}
This is a physically interesting observable that depends only on the
locations of the ions and not directly on the electric field.  The
correlations present in the electric field will
affect the correlations of the particle-particle energy.  All runs are
composed of 5,000 Monte Carlo equilibration steps followed by 400,000
measurement steps.  The autocorrelation functions of the
particle-particle energy for the different field update methods are
shown in Fig.~\ref{fig:ppe_autocorr}.  Integrating the autocorrelation
function out to 700, we obtain an autocorrelation time of $89.85$ for
the local heat bath method simulation, $78.10$ for the worm update
method, and $47.30$ for the FFT based update method.


\section{Conclusions}

Efficient simulation of physical systems using Monte Carlo simulation
depends on Monte Carlo update moves that decorrelate the system
rapidly.  We have studied the correlations of various observables in
simulations of charged particles using the method introduced by Maggs
et al., and were able to reduce the autocorrelation time of many
observables by introducing a method based on the FFT that produces a
completely independent electric field (subject to the Gauss' law
constraint) after a single update step.  We have shown on a physically
realistic system that, although local heat bath methods and worm-based
methods are able to decorrelate the electric field well on different
length scales, the FFT based method performs better than either of
them on all observables studied.  Additionally, using our
implementations, the FFT method took the least amount of computer
time.

\begin{acknowledgments}

The work of A.~Duncan was
supported in part by NSF grant PHY0244599. The work of R.D.~Sedgewick
and R.D.~Coalson was supported by NSF grant CHE0092285.  A.~Duncan is
grateful for the support and hospitality of the Max Planck Institute
(Heisenburg Institut f\"ur Physik und Astrophysik), where part of
this work was performed.
\end{acknowledgments}


\begin{thebibliography}{99}
\bibitem{maggs:prl} A.C.~Maggs and V.~Rossetto, Phys.\ Rev.\ Lett.\ \textbf{88}
, 196402 (2002)
\bibitem{rob1}  J. Israelachvili, \textit{Intermolecular and Surface
  Forces, 2nd Ed.} (Academic, London, 1992)
\bibitem{rob2} D. Bashford and M. Karplus, Biochem.\ \textbf{29}, 10219
  (1990)
\bibitem{rob3} D. Bashford and D.A. Case,  Biochem.\ \textbf{32}, 8045
  (1993)
\bibitem{rob4} P.G. Arscott, A.-Z. Li, and V.A. Bloomfield,
  Biopoly. \textbf{30}, 619 (1990)
\bibitem{rob5} D.A. McQuarrie, \textit{Statistical Mechanics} (Harper
  and Row, New York, 1976)
\bibitem{rob6} D. Chandler, \textit{Introduction to Modern Statistical
  Mechanics} (Oxford University Press, New York, 1987)
\bibitem{rob7} M. Zacharias, B.A. Luty, M.E. Davis, and J.A. McCammon, 
J.\ Mol.\ Biol.\ textbf{238}, 455 (1994)
\bibitem{rob8} A.K. Sood, in \textit{Solid State Physics}, edited by
  H.~Eherenreich and D.~Trunbull (Academic, New York, 1991), Vol.~45, p.~1
\bibitem{rob9} N. Ise and H. Yoshida, Acc.\ Chem.\ Res.\ \textbf{29}, 3 (1996)
\bibitem{rob10} B. Hille, \textit{Ion Channels of Excitable
  Membranes. 3rd Ed.} (Sinauer Assoc., Sunderland, 2001)
\bibitem{ewalda} J.V.L.~Beckers, C.P.~Lowe and S.W.~de~Leeuw,
  Molecular Simulation \textbf{20}, 269 (1988)
\bibitem{ewaldb} J.W.~Perram, H.G.~Petersen and S.W.~de~Leeuw,
  Molecular Phys.\ \textbf{65}, 875 (1985)
\bibitem{fft} E.~Esssmann, L.~Perera, M.L.~Berkowitz, T.~Darden,
  H.~Lee and L.G.~Pedersen, J.\ Chem.\ Phys.\ \textbf{103}, 8577 (1995)
\bibitem{ourpaper} A.~Duncan, R.D.~Sedgewick, and R.D.~Coalson, Phys.\ Rev.\ E \textbf{71}, 046702 (2005)
\bibitem{maggs:worm} L.~Levrel, F.~Alet, J.~Rottler, and
A.C.~Maggs, Statphys22 Proceedings, to be published in PRAMANA
[also available at  \texttt{cond-mat/0409350}]
\bibitem{maggs:trail} L.~Levrel and A.C.~Maggs, \texttt{cond-mat/0503744} (2005)
\bibitem{alet:orig_worm} F.~Alet and E.~S\o rensen, Phys.\ Rev.\ E
  \textbf{67}, 015701 (2003)
\bibitem{tonyqcd} A.~Duncan, E.~Eichten and H.~Thacker,
  Phys.\ Rev.\ Lett.\ \textbf{76}, 3894 (1996)
\bibitem{fftw} M.~Frigo and S.G.~Johnson, Proc.\ of the IEEE \textbf{93}, 216-231 (2005)
\end{thebibliography}
\end{document}